\documentclass[journal]{IEEEtran}
\usepackage[latin9]{inputenc}
\usepackage[pdftex]{graphicx}
\usepackage{hyperref, amsmath, cite}

\begin{document}
% paper title% can use linebreaks \\ within to get better formatting as desired

\title{Quantum optical technologies\\ for metrology, sensing and imaging}

\author{\IEEEauthorblockN{Jonathan P. Dowling, \emph{Fellow, OSA},\thanks{J. P. Dowling is Professor and Hearne Chair of Theoretical Physics at the Department of Physics and Astronomy, and Co-Director of the Hearne Institute for Theoretical Physics, at Louisiana State University. (Email: jdowling@lsu.edu)} and Kaushik P. Seshadreesan, \emph{Student Member, OSA}, \thanks{K. P. Seshadreesan is a PhD student of Physics at Louisiana State University. (Email: ksesha1@lsu.edu)}\\}
\IEEEauthorblockA{Hearne Institute for Theoretical Physics and Department of Physics and Astronomy, Louisiana State University, Baton Rouge, LA 70803, USA\\}}

%\author{
%	\IEEEauthorblockN{
%	Christian Häger, \emph{Student Member, IEEE},
%	Alexandre Graell i Amat, \emph{Senior Member, IEEE},
%	Fredrik Brännström, \emph{Member, IEEE},
%	Alex Alvarado, \emph{Member, IEEE}, and
%	Erik Agrell, \emph{Senior Member, IEEE}
%	\thanks{Parts of this paper have been presented at the
%	European Conference on Optical Communication (ECOC), Cannes, France,
%	Sep. 2014.}
%	\thanks{This work was partially funded by the Swedish Research Council under
%	grant \#2011-5961 and by the 
%  Engineering and Physical Sciences Research Council (EPSRC) project UNLOC (EP/J017582/1), United Kingdom. The
%	simulations were performed in part on resources provided by the
%	Swedish National Infrastructure for Computing (SNIC) at C3SE. } 
%	\thanks{C.~Häger, A.~Graell i Amat, F.~Brännström, and E.~Agrell are with the
%	Department of Signals and Systems, Chalmers University of Technology, Gothenburg,
%	Sweden (emails: \{christian.haeger, alexandre.graell,
%	fredrik.brannstrom,
%	agrell\}@chalmers.se). }
%	\thanks{A.~Alvarado is with the Optical Networks
%	Group, Department of Electronic and Electrical Engineering,
%	University College London, London WC1E7JE, UK (email:
%	alex.alvarado@ieee.org).}
%	}
%}

%\author{Kaushik~P.~Seshadreesan \thanks{Hearne Institute for Theoretical Physics and Department of Physics and Astronomy, Louisiana State University, Baton Rouge, LA 70803, USA} \and~Jonathan~P.~Dowling\thanks{Beijing Computational Science Research Center, Beijing, 100084, China}}

\maketitle
\begin{abstract}
Over the past 20 years, bright sources of entangled photons have led
to a renaissance in quantum optical interferometry. Optical interferometry
has been used to test the foundations of quantum mechanics and implement
some of the novel ideas associated with quantum entanglement such
as quantum teleportation, quantum cryptography, quantum lithography,
quantum computing logic gates, and quantum metrology. In this paper,
we focus on the new ways that have been developed to exploit quantum
optical entanglement in quantum metrology to beat the shot-noise limit,
which can be used, e.g., in fiber optical gyroscopes and in sensors
for biological or chemical targets. We also discuss how this entanglement
can be used to beat the Rayleigh diffraction limit in imaging systems
such as in LIDAR and optical lithography. 
\end{abstract}
% IEEEtran.cls defaults to using nonbold math in the Abstract.% This preserves the distinction between vectors and scalars. However,% if the journal you are submitting to favors bold math in the abstract,% then you can use LaTeX's standard command \boldmath at the very start% of the abstract to achieve this. Many IEEE journals frown on math% in the abstract anyway.

% For peer review papers, you can put extra information on the cover% page as needed:% \ifCLASSOPTIONpeerreview% \begin{center} \bfseries EDICS Category: 3-BBND \end{center}% \fi% For peerreview papers, this IEEEtran command inserts a page break and% creates the second title. It will be ignored for other modes.

\IEEEpeerreviewmaketitle{}

\section{Introduction}

Sensors are an integral part of many modern technologies that touch
our day to day lives, e.g., automotive technologies, the global
positioning system (GPS) and mobile telecommunication, to name a few. They also get widely used in industrial applications,
e.g, in manufacturing and machinery, in petroleum-well mapping, oil refineries, chemical processes and medicine. 
%Optical sensors, among others, find wide use especially in imaging. For example, phase imaging is used in microscopy to track and study particles at micro- and nano-scales in real time~\cite{TJDK13}.
It is desired for a sensor to capture the faintest of signals. The capability of a sensor to do so is largely dictated by its noise or precision characteristics. Hence, metrology---the study of precision measurements---plays a fundamental role in the design of sensors. 

Quantum mechanics, being a fundamental theory of nature, has a bearing
on the performance of technologies that are based on information processing, e.g., computation, communication and
cryptography. It is thus imperative to consider quantum
mechanics in order to determine the ultimate limits of these technologies.
To this end, the classical theories of computation, communication
and cryptography---which revolutionized the technological world in the
past half a century or so---have been revisited to study the
effects of quantum mechanics. This has lead to exciting new possibilities
in these areas, such as quantum algorithms for fast integer factorization,
fast database search, quantum teleportation, superdense coding and
quantum key distribution~\cite{Shor97, Grov96, BBCJPW93, Eke91, BB84}. Likewise, metrology, which is also a science based on information acquisition, processing, and estimation, has been revisited to include the effects of quantum
mechanics too. Quantum metrology~\cite{GLM06} has been found to enable measurements with precisions that surpass the classical limit, and has grown into an exciting new area of research with potential applications, e.g., in gravitational wave detection~\cite{PRRH11},
quantum positioning and clock synchronization~\cite{GLM01}, quantum frequency standards~\cite{HMPEP97},
quantum sensing~\cite{WVBGN13, CLMPN12}, quantum radar and LIDAR~\cite{JLGD13, GAWLLD10}, quantum imaging~\cite{Sh07, LGB02} and quantum
lithography~\cite{KBD04,KBAWBD01, BKABCD00}.

\begin{figure}[hbt] 
\begin{center}%Frame position
\includegraphics[scale=0.6]{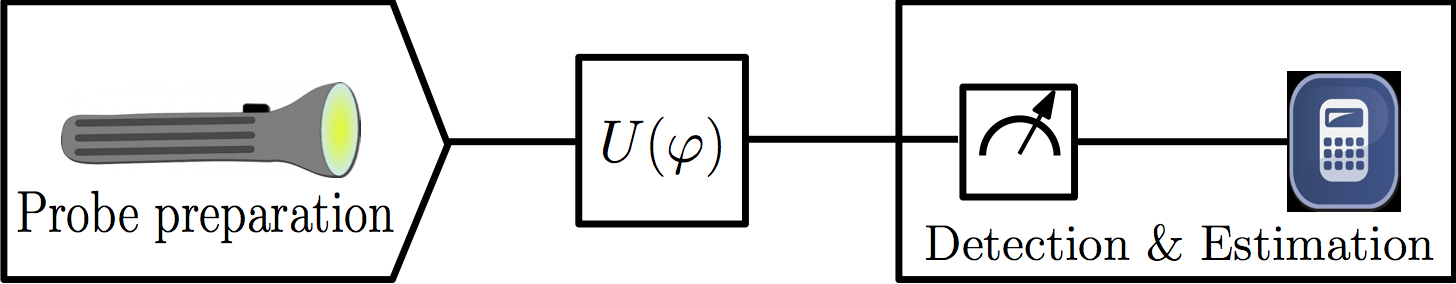}
\end{center}
\caption{A schematic of a typical quantum parameter estimation setup. Probes prepared in suitable quantum states are made to evolve through a unitary process $U(\varphi)$, which is an optical interferometer in our case. The process imparts information about the unknown parameter of interest on to the probes. The probes are then detected at the output, and the measurement outcomes used to estimate the unknown value of the parameter.}
\label{paraest_intro}
\end{figure}

Quantum metrology offers a theoretical framework that can be used to analyze
the precision performance of measurement devices that employ quantum-mechanical
probes containing nonclassical effects such as entanglement or squeezing.
It relies on the theory of quantum parameter estimation~\cite{He76, BC94, BCM96}.
Consider the typical scenario of parameter estimation described in
Fig.~\ref{paraest_intro}, where we want to estimate an unknown parameter associated with the unitary dynamics generated by a known physical
process. We prepare probes in suitable quantum states, evolve them
through the process, and measure the probes at the output using a
suitable detection strategy. We then compare the input and output
probe states, which allows us to estimate the unknown parameter of
the physical process. Let us suppose that the generating Hamiltonian is linear in the number of probes. When $N$ classical probes (probes with
no quantum effects) are used, the precision is limited by a scaling
given by $1/\sqrt{N}$; known as the shot-noise limit. This scaling arises from the central limit theorem of
statistics. On the other hand, probes with quantum entanglement can
reach below the shot-noise limit and determine the unknown parameter
with a precision that can scale as $1/N$; known as the Heisenberg
limit~\cite{HB93}.

\begin{figure}[hbt] 
\begin{center}%Frame position
\includegraphics[scale=0.7]{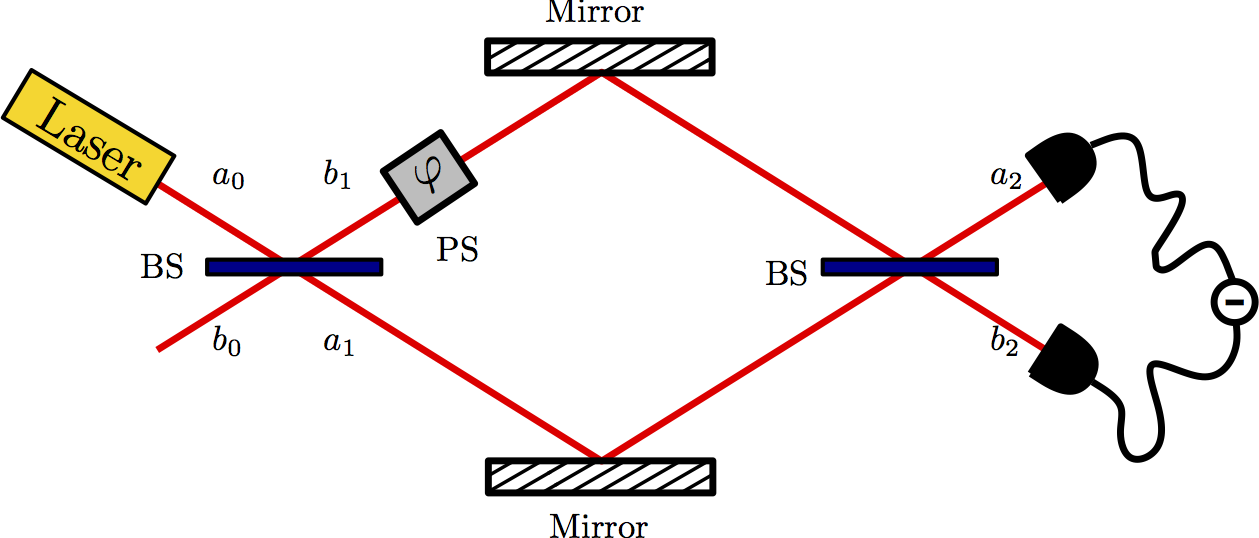}
\end{center}
\caption{A schematic of the conventional Mach-Zehnder interferometry based on coherent light input and intensity difference detection. The BS and PS denote beamsplitters and phase shifters.}
\label{mzi_intro}
\end{figure}

Optical metrology uses light interferometry as a tool to perform precision
measurements. The most basic optical interferometer is a two-mode device with an unknown relative phase (between the two modes). This unknown phase can be engineered to carry information about different quantities of interest in different
contexts, e.g., it is related to the strength of a magnetic field in an optical magnetometer,
a gravitational wave at LIGO (light interferometer gravitational wave
observation), etc. Fig.~\ref{mzi_intro} shows a conventional optical interferometer in the Mach Zehnder configuration. The input to the classical interferometer is a coherent laser source, and the detection is based on intensity difference measurement.
When a coherent light of average photon number $\overline{n}$ is used, the precision of phase estimation is limited
by the shot-noise of $1/\sqrt{\overline{n}}$ associated
with the intensity fluctuations at the output, which have their origin
in the vacuum fluctuations of the quantized electromagnetic field that enter the device through the unused input port $b_0$.

However, quantum optical metrology enables
sub-shot-noise phase estimation. In a seminal work in the field, Caves~\cite{Ca81} showed that when the nonclassical squeezed vacuum state is mixed instead of the vacuum state in the
unused port of the same interferometer, sub-shotnoise precision that
scales as $1/\overline{n}^{2/3}$ can be attained.
Subsequently, two-mode squeezed states were shown to enable phase
estimation at a precision of $1/\overline{n}$~\cite{BS84}. With the advancement in single-photon technology, finite photon number states containing quantum entanglement were also proposed and studied in quantum optical metrology. This includes the $N00N$ states~\cite{Do08}, which are Schr{\"o}dinger cat-like, maximally mode-entangled states of two modes, where the $N$ photons are in superposition of all $N$ photons being in one mode or the other; the Holland-Burnett states~\cite{HB93}and the Berry-Wiseman states~\cite{BW00}, to name a few. All these states were found to be capable of attaining the Heisenberg limit $1/N$. The above theoretical results have led to many experimental demonstrations of sub-shot-noise metrology with finite photon number states~\cite{VDGJK14, XHBWP11, AAS10, HBBWP07, NOOST07, Bo04}.
%Recently,
%in a revival of the study on indefinite photon number states, the interferometry with coherent state mixed with squeezed
%vacuum state has been revisited, and the case when the coherent state
%and squeezed vacuum states are mixed in equal intensities has been
%shown to be capable of phase precision that scales as $1/\langle\bar{N}\rangle$.
%More indefinite photon number states such as the entanglement coherent
%states, the even and odd cat states, etc, have also been shown to
%be capable of $1/\langle\bar{N}\rangle$ precision.

Along with the different quantum states of light, a plethora of detection
strategies have also been investigated. This includes homodyne and
heterodyne detection~\cite{YC83}, the canonical phase measurement~\cite{PB88} (which can be mimicked by an adaptive
measurement~\cite{Wise95}),
photon number counting~\cite{RLMN05, KYS08}, and photon number parity~\cite{GM10}. These measurement
schemes have been shown to be capable of attaining the optimal precisions
of different quantum states of light. 

More recently, numerous
studies have been devoted to investigating the effects of photon loss, dephasing noise and other decoherence phenomena, on the
precision of phase estimation in quantum optical metrology. Useful lower bounds on precision, and
optimal quantum states that attain those bounds, have been identified in some such scenarios both
numerically and analytically~\cite{KCD14, DKM12, EFD11, KSD11, LHLKMM09, DBSLWBW09}.

In this paper, we focus on interferometry with entangled states of finite photon number, in particular, those based on the $N00N$ states. We focus on some recent
experiments that demonstrate $N00N$ state
interferometry for modest photon numbers $N$, with potential applications in quantum technology~\cite{IRS14, RBMOF13, OOT13, WVBGN13, CLMPN12}. The $N00N$ states not only attain the Heisenberg limit of $1/N$ in phase
precision---known as super-sensitivity, but are also capable of phase resolution below the Rayleigh
diffraction limit---known as super-resolution~\cite{Do08}. The paper is organized as follows. In section~\ref{bas_con}, we discuss some basic concepts of quantum optical metrology. This includes the different available representations to study two-mode quantum interferometry, and the methods of quantum parameter estimation theory that are used to analyze the interferometric output statistics to estimate the unknown phase. In section~\ref{qom}, we discuss quantum optical metrology. We begin by introducing quantum entanglement, which is the driving-force behind the quantum enhancement, and describe the different available methods to generate entanglement for optical metrology. We describe an interferometric scheme that is known to achieve the optimal Heisenberg limit in phase precision. Section~\ref{qt} is devoted to quantum technologies for metrology, sensing and imaging. In this section, we present results from a few recent experiments that have demonstrated the benefits of quantum optical interferometry with the $N00N$ states for such technological applications. In Section~\ref{disc}, we conclude with a brief discussion on some recent considerations in quantum optical interferometry and a summary.

\section{Basic Concepts}
\label{bas_con}

We now discuss the basic concepts that underlie quantum optical metrology, namely, those of quantum optical interferometry and quantum parameter estimation.

\subsection{Classical optical interferometry}
Before we describe quantum interferometry, let us briefly examine the conventional coherent laser light interferometer described in Fig.~\ref{mzi_intro} in purely classical terms. The input laser beam is split into two beams of equal intensities by the first 50:50 beamsplitter. These beams then gather an unknown relative phase as they pass through the device. They are then recombined on the final beamsplitter, and the average intensity difference between the two output beams is measured. A simple classical optics calculation tells us that the intensities at the output ports may be written in terms of the input intensity $I_{a_0}$ and the relative phase $\varphi$ as
\begin{align}
I_{a_2}&=I_{a_0}\sin^2(\varphi/2),\nonumber\\
I_{b_2}&=I_{a_0}\cos^2(\varphi/2).
\end{align}
This implies the intensity difference between the two output ports  is $M(\varphi)\equiv I_{b_2}-I_{a_2}=I_{a_0}\cos\varphi$---sinusoidal fringes that can be observed when the relative phase is varied. 

The precision with which one can estimate an unknown relative phase based on the measurement of $M$, in terms of the phase error, or the minimum detectable phase, $\Delta\varphi$, may be determined to a good approximation using the following linear error-propagation formula:
\begin{align}
\Delta\varphi=\frac{\Delta M}{\left|d M/d \varphi\right|}=\frac{\Delta M}{I_{a_0}\sin\varphi}.
\end{align}
The above equation suggests that at a local value of phase $\varphi=\pi/2$, the precision of phase estimation can be made arbitrarily small by measuring the intensity difference $M$ with infinite precision, and further by making the input intensity $I_a$ arbitrarily large. However, quantum mechanics rules out the possibility of measuring intensities with infinite precision, i.e., with $\Delta M=0$. This is because photon detection is intrinsically a quantum phenomenon, where what is actually measured is not a continuously varying intensity signal, but rather the discrete number of quanta of energy, or photons, that are absorbed by the detector. This absorption process is inherently stochastic due to the vacuum fluctuations of the quantized electromagnetic field, and in the case of the coherent laser light the photon numbers detected obey a Poisson distribution. This limits the precision of phase estimation in classical interferometry to $\Delta\varphi=1/\sqrt{\overline{n}}$, where $\overline{n}$ is the intensity of the input laser beam.

\subsection{Quantum optical interferometry}

In order to give a fully quantum treatment of two-mode optical interferometry, we now introduce a quantized mode of the electromagnetic field and describe its various states. We then discuss the most relevant linear optical transformations for a pair of independent photonic modes in interferometry, namely the beam splitter and phase shifter transformations. This is followed by considering Hermitian operators as measurement observables at the output of the interferometer, and the calculation of precision of phase estimation based on the error propagation formula.
\\
\subsubsection{Quantum states of a single mode of the electromagnetic field}

A quantized mode of the electromagnetic field is completely described by its creation and annihilation operators, $\hat{a}$ and $\hat{a}^\dagger$, which satisfy the commutation relation $\left[\hat{a},\hat{a}^\dagger\right]=1$. They are defined by their action on the number states of the mode, $|n\rangle$---also called Fock states, given by:
\begin{align}
\hat{a}|n\rangle=\sqrt{n}|n-1\rangle, \ \hat{a}^\dagger|n\rangle=\sqrt{n+1}|n+1\rangle.
\end{align}
Pure states of the single-mode field (vectors in Hilbert space) can be expressed in terms of the action of a suitable function of the mode creation and annihilation operators on the vacuum state $|0\rangle$. For example, a Fock state $|n\rangle$ can be expressed as $\frac{\hat{a}^\dagger}{\sqrt{n!}}|0\rangle$, where $|0\rangle$ is the vacuum state. The coherent state can be written as:
\begin{equation}
|\alpha\rangle=\hat{D}(\alpha)|0\rangle,
\end{equation}
where $\hat{D}(\alpha)=\exp\left(\alpha \hat{a}^\dagger-\alpha^*\hat{a}\right)$ is called the displacement operator, and $\alpha$ is a complex number that denotes the coherent amplitude of the state. Both the Fock states and the Coherent states form complete bases (the coherent states in fact form an over-complete basis). Therefore, any pure state of the quantum single-mode field can be expressed in terms of these states. More generally, any state of the single-mode field, including mixed states, which are ensembles of pure states, can be written in terms of these states in the form of a density operator. Density operators are positive semi-definite trace one operators. The positive semi-definiteness condition enforces that the eigenvalues of the state are non-negative real numbers, so that they can be interpreted as valid probabilities. The trace one condition further ensures that these probabilities sum up to one, thereby making the state a normalized state. For example, the most general state of a single-mode field can be written in the Fock basis as the following density operator:
\begin{equation}
\label{densop}
\hat{\rho}=\sum_{n,n'}p_{n,n'}|n\rangle\langle n'|, \ \operatorname{Tr}(\rho)=1, \ \rho\geq0.
\end{equation}

Alternatively, a quantized mode can be described in terms of quasi-probability distributions in the phase space of eigenvalues $x$ and $p$ of the quadrature operators of the mode $\hat{x}$ and $\hat{p}$. These operators are defined in terms of the creation and annihilation operators of the mode as $\hat{x}=\hat{a}^\dagger+\hat{a}$ and $\hat{p}=i(\hat{a}^\dagger-\hat{a})$, respectively. The Wigner distribution of a single-mode state can be obtained from its density operator of the form in (\ref{densop}) as:
\begin{equation}
W(\alpha)=\frac{1}{2\pi^2}\int d^2\tilde{\alpha} \operatorname{Tr}\left\{\hat{\rho} \hat{D}(\tilde{\alpha})\right\}e^{-\tilde{\alpha}\alpha^*-\tilde{\alpha}^*\alpha},
\end{equation}
where $\tilde{\alpha}=\tilde{x}+i\tilde{p}$ and $\alpha=x+ip$.
\\
\subsubsection{Quantum states and dynamics in the Mach-Zehnder interferometer}
In the quantum description of the Mach-Zehnder interferometer (MZI), we associate mode creation and annihilation operators with each of the two modes. Here, we call them $\hat{a}_i$, $\hat{a}^\dagger_i$ and $\hat{b}_i$, $\hat{b}^\dagger_i$, $i\in \{0, 1,2\}$, where the different values of $i$ refer to the modes at the input, inside, and output of the interferometer. The two modes of an MZI could be spatial modes or polarization modes.

Consider the propagation of the input quantum states of the two modes through the different linear optical elements present in the MZI. In the so-called Heisenberg picture, the propagation can be viewed as a transformation of the mode operators via a scattering matrix $M_i$:
\begin{equation}
\left[
\begin{array}{c}
\hat{a}_0\\
\hat{b}_0
\end{array}
\right]=\hat{M_i}^{-1}
\left[
\begin{array}{c}
\hat{a}_1\\
\hat{b}_1
\end{array}
\right].
\label{eq:opTRANS}
\end{equation}
The scattering matrices corresponding to a 50:50 beam splitter and a phase shifter are given by
\begin{eqnarray}
\hat{M}_{\rm BS}=\frac{1}{\sqrt{2}}  \left[
\begin{array}{lr}
1   & i \\
i & 1
\end{array}
\right] ,\
\label{eq:UxBS}
%\text{and}
\hat{M}_\varphi=\left[
\begin{array}{lr}
1    & 0 \\
0 & e^{-i\varphi}
\end{array}
\right],
\label{eq:UxPh}
\end{eqnarray}
respectively.
(Note that this form for $\hat{M}_{\rm BS}$ holds for beamsplitters that are constructed as a single dielectric layer, in which case the reflected and the transmitted beams gather a relative phase of $\pi/2$.) The two-mode quantum state at the output of a MZI in the Fock basis, can therefore be obtained by replacing the mode operators in the input state in terms of the output mode operators, where the overall scattering matrix is given by:
$\hat{M}_{\rm
  MZI}=\hat{M}_{\rm BS}\hat{M}_\varphi\hat{M}_{\rm BS}$ and is found to be:
\begin{equation}
\hat{M}_{\rm MZI}=
i e^{-i \frac{\varphi}{2}}\left[
\begin{array}{cc}
\sin\frac{\varphi}{2} & \cos\frac{\varphi}{2}\\
 \cos\frac{\varphi}{2}& -\sin\frac{\varphi}{2}
\end{array}
\right].
\label{scatscat}
\end{equation}
(Note that the overall scattering matrix has been suitably renormalized.)

In terms of phase space quasi-probability distributions such as the Wigner distribution function, the propagation through the Mach-Zehnder interferometer can be similarly described by relating the initial complex variables in the
Wigner function to their final expressions as: 
\begin{equation}
W_{\rm out}(\alpha_1,\beta_1)=W_{\rm in}\left[\alpha_0(\alpha_1,\beta_1),\beta_0(\alpha_1,\beta_1)\right].
\label{eq:WafterBS}
\end{equation}
The relation between the complex variables is similarly given in terms of the two-by-two scattering matrices $\hat{M}$: 
\begin{equation}
\left[
\begin{array}{c}
\alpha_0\\
\beta_0
\end{array}
\right]=\hat{M}^{-1}
\left[
\begin{array}{c}
\alpha_1\\
\beta_1
\end{array}
\right],
\label{eq:AMPTRANS}
\end{equation}
$\alpha_0$, $\beta_0$, $\alpha_1$, and $\beta_1$ being the complex
amplitudes of the field in the modes $\hat{a}_0$, $\hat{b}_0$, $\hat{a}_1$,
and $\hat{b}_1$, respectively. The approach based on phase space probability distributions is particularly convenient and powerful when dealing with Gaussian states, namely, states that have a Gaussian Wigner distribution, and Gaussian operations~\cite{WPGNRSL12}. Examples include the coherent state, the squeezed vacuum state and the thermal state~\cite{GK05}. This is due to the fact that a Gaussian distribution is completely described by its first and second moments, and there exist tools based on the algebra of the symplectic group that can be used to propagate the mean and covariances of Gaussian states of any number of independent photonics modes.
\\
\subsubsection{The Schwinger model}

The Schwinger model presents an alternative way to describe quantum states and their dynamics in a MZI~\cite{YMK86}. The model is based on an interesting relationship between the algebra of the mode operators of two independent photonic modes and the algebra of angular momentum. 

Consider the following functions of the mode operators of a pair of independent photonic modes, say, $\hat{a}_1$, $\hat{a}_1^{\dagger}$, $\hat{b}_1$, and $\hat{b}_1^{\dagger}$:
\begin{eqnarray}
\label{shhhh}
&\hat{J}_x=\frac{1}{2}(\hat{a}_1^{\dagger}\hat{b}_1+\hat{b}_1^{\dagger}\hat{a}_1),\ \hat{J}_y=\frac{1}{2 i}(\hat{a}_1^{\dagger}\hat{b}_1-\hat{b}_1^{\dagger}\hat{a}_1),&\nonumber\\
&\hat{J}_z=\frac{1}{2}(\hat{a}_1^{\dagger}\hat{a}_1-\hat{b}_1^{\dagger}\hat{b}_1).&
\end{eqnarray}
(Note that we have chosen the mode index ``1", which in the MZI of Fig.~\ref{mzi_intro} corresponds to the modes past the first beamsplitter. This is relevant when we discuss operational usefulness of the $\hat{J}_q$ operators shortly.) They can be shown to obey the SU(2) algebra of angular momentum operators, namely, $\left[\hat{J}_q, \hat{J}_r\right]=i\hbar\epsilon_{q,r,s}\hat{J}_s$, where $\epsilon$ is the antisymmetric tensor and where $q,r,s \in\{x,y,z\}$. Based on this relation, a two-mode $N$-photon pure state gets uniquely mapped on to a pure state in the spin-$N/2$ subspace of the angular momentum Hilbert space, i.e.,
\begin{align}
|n_a, n_b\rangle\rightarrow |j=\frac{n_a+n_b}{2}, m=\frac{n_a-n_b}{2}\rangle.
\end{align}
The propagation of the quantized single-mode field can be viewed in terms of the Schwinger representation as a SU(2) group transformation generated by the angular momentum operators $\hat{J}_x$, $\hat{J}_y$ and $\hat{J}_z$. For example, the beamsplitter transformation of (\ref{eq:UxBS}) can be written as:
\begin{equation}
\left[
\begin{array}{c}
\hat{a}_0\\
\hat{b}_0
\end{array}
\right]=U_{\rm BS}^\dagger
\left[
\begin{array}{c}
\hat{a}_1\\
\hat{b}_1
\end{array}
\right]U_{\rm BS},
\label{eq:bsuTRANS}
\end{equation}
where $U_{\rm BS}=\exp(i\pi/2\hat{J}_x)$, and likewise, the transformation due to the phase shifter inside the interferometer can be described as
\begin{align}
\left[
\begin{array}{c}
\hat{a}_1\\
\hat{b}_1
\end{array}
\right]
\rightarrow U_{\varphi}^\dagger
\left[
\begin{array}{c}
\hat{a}_1\\
\hat{b}_1
\end{array}
\right]U_\varphi,
\label{eq:psuTRANS}
\end{align} 
Using the SU(2) algebra of the angular momentum operators and the Baker-Hausdorff lemma~\cite{sakurai}, the overall unitary transformation corresponding to the MZI can be expressed as $\hat{U}_{MZI}=\textrm{exp}(-i\varphi \hat{J}_y)$. Operationally, for any given two-mode state, the operator $\hat{J}_z$ tracks the photon number difference between the two modes inside the interferometer (which is $\propto\hat{a}_1^{\dagger}\hat{a}_1-\hat{b}_1^{\dagger}\hat{b}_1$). Similarly, it can be shown using the SU(2) commutation relations that the operators $\hat{J}_x$ and $\hat{J}_y$ track the photon number differences at the input (which is $\propto\hat{a}_0^{\dagger}\hat{a}_0-\hat{b}_0^{\dagger}\hat{b}_0$) and the output (which is $\propto\hat{a}_2^{\dagger}\hat{a}_2-\hat{b}_2^{\dagger}\hat{b}_2$), respectively.
\\
\subsubsection{Measurement and phase estimation}
After propagating the two-mode quantum state through the MZI, we measure the output state (most generally a density operator $\hat{\rho}$) using a suitable Hermitian operator $\hat{O}$ as the measurement observable. For example, the measurement observable corresponding to the intensity difference detection of the conventional MZI described in Fig.~\ref{mzi_intro} is the photon number difference operator $\hat{O}=\hat{b}_2^\dagger\hat{b}_2-\hat{a}_2^\dagger\hat{a}_2$. Another interesting detection scheme that has been found to be optimal for many input states is the photon number parity operator~\cite{SKDL13, SALD11, ARCPHLD10, GM10} of one of the two output modes, e.g., the parity operator of mode $\hat{a}_2$ is given by $\hat{\Pi}=(-1)^{\hat{a}_2^\dagger \hat{a}_2}$. The measured signal corresponding to any observable $\hat{O}$ is given by $\langle\hat{O}\rangle=\operatorname{Tr}\{\hat{O}\hat{\rho}\}$. Further, the precision with which the unknown phase $\varphi$ can be estimated using the chosen detection scheme, to a good approximation, is given using the error propagation formula as
\begin{align}
\label{errprop}
\Delta\varphi=\frac{\Delta O}{\left|d \langle\hat{O}\rangle/d \varphi\right|},
\end{align}
where $\Delta O=\sqrt{\langle\hat{O}^2\rangle-\langle\hat{O}\rangle^2}$.

As an example, consider the coherent light interferometer of Fig.~(\ref{mzi_intro}). The output state is determined using the scattering matrix of (\ref{scatscat}) as
\begin{multline}
|\alpha\rangle|0\rangle\rightarrow|i\alpha\sin(\varphi/2) e^{-i\varphi/2}\rangle|i\alpha\cos(\varphi/2) e^{-i\varphi/2}\rangle.
\end{multline}
The output signal for the measurement operator $\hat{O}=\hat{b}_2^\dagger\hat{b}_2-\hat{a}_2^\dagger\hat{a}_2$ corresponding to intensity difference detection is 
\begin{equation}
\langle\hat{O}\rangle=|\alpha|^2(\cos^2(\varphi/2)-\sin^2(\varphi/2))=|\alpha|^2\cos\varphi,
\end{equation}
which matches the classical result. The second moment $\langle\hat{O}^2\rangle$ for the output state is $|\alpha|^4\cos^2\varphi+|\alpha|^2$, with which we can then ascertain the precision of phase estimation possible with the coherent light interferometer and intensity difference measurement to be
\begin{align}
\Delta\varphi&=\frac{\sqrt{|\alpha|^4\cos^2\varphi+|\alpha|^2-|\alpha|^4\cos^2\varphi}}{\left|\alpha\right|^2\sin\varphi}\nonumber\\
&=\frac{1}{|\alpha||\sin\varphi|}=\frac{1}{\sqrt{\overline{n}}|\sin\varphi|},
\end{align}
where $\overline{n}$ is the average photon number of the coherent state. Say the unknown phase $\varphi$ is such that $\varphi-\theta$ is a small real number, where $\theta$ is a control phase. Then, the precision is optimal when $\theta$ is an odd multiple of $\pi/2$, attaining $\Delta\varphi=1/\sqrt{\overline{n}}$, which is the quantum shot-noise limit. 

In the above, it is possible to get rid of the dependence on the actual value of phase by considering the fringe visibility observable~\cite{Eng96}. The visibility observable accomplishes this by keeping track of not only the photon number difference, but also the total photon number observed.

%The photon number difference observable, in the Schwinger model, becomes the $\hat{J}_y$ operator given in (\ref{shhhh}). 

\subsection{Quantum parameter estimation}

Although the linear error propagation formula described in (\ref{errprop}) provides a good approximation for the precision of estimation of the unknown phase $\varphi$ of an optical interferometer, in order to determine the absolute lower bound on precision that is possible in a given interferometric scheme, one has to resort to the theory of parameter estimation. We now briefly review the quantum theory of parameter estimation. There exist two main paradigms in parameter estimation, (i) where the unknown parameter is assumed to hold a deterministic value, (ii) where the unknown parameter is assumed to be intrinsically random. Here, we focus on the first one.

Consider $N$ identical copies of a quantum state that has interacted with the unknown parameter of interest and holds information about it. Since the state carries the information about the parameter of interest, say $\varphi$, let us denote it as $\hat{\rho}_\varphi$. Now, consider a set of data points ${\rm x}=\{x_1,x_2,\dots,x_\nu\}$ that are obtained from the $N$ copies of $\hat{\rho}_\varphi$ as outcomes of a generalized quantum measurement. The generalized measurement is a positive operator-valued measure (POVM), which is a collection of positive operators $\Lambda_\mu$, with the index $\mu\in\{1,2,\dots,M\}$ denoting the outcome of the measurement, whose probability of occurrence for a state $\rho$, is given by $p(\mu)=\operatorname{Tr}\{\rho \Lambda_x\}$. The elements of a POVM add up to the identity $\sum_x\Lambda_x=I$, which ensures that $p(\mu)$ is a valid probability distribution. Since the data points are obtained by measuring identical copies of the quantum state , the $x_i$s are realizations of independent and identically distributed random variables $X_i$, $i\in\{1,2,\dots,\nu\}$ that are distributed according to some probability distribution function $p_\varphi(X)$. The goal is to apply a suitable estimation rule $\widehat{\varphi}_\nu$ to the data points, to obtain a good estimate for the unknown parameter $\varphi$.

When estimation rule $\widehat{\varphi}_\nu$ is applied to a set of data points $x$, a good measure of precision for the resulting estimate $\widehat{\varphi}_\nu({\rm x})$ is its mean-square error, given by:
\begin{equation}
\Delta^2\widehat{\varphi}_\nu=\operatorname{E}[(\widehat{\varphi}_\nu({\rm x})-\varphi)^2],
\end{equation}
where $\operatorname{E}$ denotes expectation value. For any estimation rule $\widehat{\varphi}_\nu$, which is unbiased, i.e.,
\begin{equation}
\operatorname{E}[\widehat{\varphi}_\nu({\rm x})]=\varphi,
\end{equation}
the Cramer-Rao theorem of classical estimation theory lower bounds the mean-square error as
\begin{equation}
\Delta^2\widehat{\varphi}_\nu\geq\frac{1}{\nu F(p_\varphi)},
\end{equation}
where $F(p_\varphi)$ is known as the Fisher information of the probability distribution given by
\begin{equation}
F(p_\varphi)=F_{\rm Cl}(\hat{\rho}_\varphi, \Lambda_\mu)=\operatorname{E}\left[-\frac{d^2}{d\varphi^2}\log p_\varphi\right].
\end{equation}
The above lower bound is called the {\it classical Cramer-Rao bound}. It gives the optimal precision of estimation that is possible when both the parameter-dependent quantum state and the measurement scheme are specified. Estimation rules that attain the classical Cramer-Rao bound are called efficient estimators. The maximum-likelihood estimator is an example of an efficient estimator that attains the lower bound in the asymptotic limit. 

The quantum theory of parameter estimation further provides an ultimate lower bound on precision of estimation when the quantum state alone is specified. It goes by the name of {\it quantum Cramer-Rao bound}, and is given by
\begin{equation}
\Delta^2\widehat{\varphi}_\nu\geq\frac{1}{\nu F_Q(\hat{\rho}_\varphi)},
\end{equation}
where $F_Q(\hat{\rho}_\varphi)$ is known as the quantum Fisher information, which is defined as the optimum of the classical Fisher information over all possible generalized measurements:
\begin{equation}
\label{qfish1}
F_Q(\hat{\rho}_\varphi)=\max_{\Lambda_\mu}F_{\rm Cl}(\hat{\rho}_\varphi, \Lambda_\mu).
\end{equation}
A measurement scheme that attains this lower bound is called an optimal measurement scheme. The symmetric logarithmic derivation operation is one such measurement, which is known to be optimal for all quantum states~\cite{BCM96}. 

In the case of entangled pure states, the quantum Fisher information takes the simplified expression given by
\[F_Q=4\Delta^2 H,\]
where $\hat{H}$ is the generator of parameter evolution. This gives rise to a generalized uncertainty relation between the generating Hamiltonian of parameter evolution and the estimator that is used for estimating the unknown value of the parameter, given by
\begin{equation}
\label{genun}
\Delta^2\widehat{\varphi}_\nu\Delta^2 H\geq\frac{1}{4\nu}
\end{equation}
for a generating Hamiltonian $\hat{H}$, and where $\nu$ is the number of data points gathered from measuring identical copies of the state. 

\section{Quantum optical metrology}
\label{qom}
Having discussed the necessary tools, we now describe quantum optical metrology. We begin the section with a brief account of the nonclassical effects that form the  source of the quantum advantage in optical metrology, namely entanglement and squeezing.

\subsection{Entanglement}

Quantum mechanics allows for correlations between physical systems beyond those allowed in classical physics. Entanglement~\cite{HHHH09} is the most prominent manifestation of such quantum correlations. Entanglement is considered by many as the hallmark feature of quantum mechanics, and is widely believed to be the source of the quantum advantage over classical techniques in quantum information processing technologies such as quantum computing, communication and cryptography. A quantum state is said to be entangled if it is anything but a separable state. For example, in the bipartite case (i.e. when there are two subsystems, say $A$ and $B$), separable states are of the form
\begin{equation}
\hat{\rho}_{AB}=\sum_x p(x)\hat{\rho}^x_A\otimes\hat{\rho}^x_B, \ \ p(x)\geq 0\ \ \forall x,\ \ \sum_x p(x)=1,
\end{equation}
where $\hat{\rho}^x_A$ and $\hat{\rho}^x_B$ are density operators. Entangled states can be made to violate a Bell's inequality, the latter being bounds on correlations possible in classical, local hidden-variable theories~\cite{BCPSW14}.

Entanglement is also thought to be the driving force behind the enhancements possible in quantum metrology over classical approaches. The quantum Fisher information of $N$ independent probes in a separable state, i.e., without quantum entanglement, cannot exceed $N$. Since this value of the quantum Fisher information corresponds to precision at the shot-noise limit based on (\ref{qfish1}), thus, separable states cannot beat the shot-noise limit. On the other hand, the quantum Fisher information of entangled states can exceed this bound. In fact it has been shown that the Fisher information of a $N$-particle state being greater than $N$ is a sufficient condition for multipartite entanglement~\cite{PS09, GLM06}. Entangled states are therefore capable of achieving sub-shot-noise precision. However, it is important to note that the presence of entanglement is a necessary, but not sufficient condition for achieving sub-shot-noise precisions. In other words, not all entangled states offer a quantum enhancement to precision metrology~\cite{HGS10}. When the generator of parameter evolution $\hat{H}$ is linear in the number of probes, according to (\ref{genun}), the quantum Fisher information of a state containing $N$ probes can at best attain a value of $N^2$, which corresponds to the Heisenberg limit in the precision of parameter estimation.

In two-mode optical interferometry, e.g., of the type in Fig.~(\ref{mzi_intro}), the relevant type of entanglement to consider is entanglement between the two modes past the first beamsplitter, namely $a_1$ and $b_1$. The most well-known mode entangled states are the $N00N$ states~\cite{Sa89, Do08}, where are defined as
\begin{equation}
|N::0\rangle_{a_1,b_1}=\frac{1}{\sqrt{2}}(|N\rangle_{a_1}|0\rangle_{b_1}+|0\rangle_{a_1}|N\rangle_{b_1}),
\end{equation}
where $a_1$ and $b_1$ denote the two modes past the first beamsplitter. The $N00N$ has a quantum Fisher information of $N^2$ and hence is capable of achieving the Heisenberg limit in phase estimation. It is known that both the photon number difference operator and the photon number parity operator are optimal for Heisenberg-limited phase estimation with the $N00N$ states~\cite{BIWH96, Do08}. Another example of finite photon number states that are known to be capable of Heisenberg-limited precision are the Holland-Burnett states $|N\rangle_{a_0}|N\rangle_{b_0}$, which result in a mode-entangled state inside the interferometer.

\subsection{Squeezed light}
In the indefinite photon number (continuous variable) regime, entanglement is intimately connected to another nonclassical effect---squeezing. Squeezed light~\cite{GK05} refers to minimum uncertainty states of light whose fluctuations with respect to one of any two orthogonal quadratures in phase space has been reduced at the expense of increased fluctuations in the other. They are described mathematically using the squeezing operator. The single-mode squeezing operator acting on a mode $\hat{a}_0$ is given by:
\begin{equation}
\label{singmosq}
\hat{S}(\xi)=\exp\left(\frac{1}{2}(\xi\hat{a}_0^{\dagger2}-\xi^*\hat{a}_0^2)\right),
\end{equation}
where $\xi=re^{i\theta}$, $r$ and $\theta$ being the squeezing parameter and squeezing angle, respectively. The squeezed vacuum state, which is the state corresponding to the action of the squeezing operator in (\ref{singmosq}) on the vacuum state is given by
\begin{align}
\label{singmosv}
|\xi\rangle&=\hat{S}(\xi)|0\rangle\nonumber\\
&=\sum_{m=0}^{\infty}\frac{(2m)!}{2^{2m}(m!)^2}\frac{\tanh^{2m} r}{\cosh r}|2m\rangle.
\end{align}
It has a mean photon number of $\overline{n}=\sinh^2r$. There are numerous ways to generate squeezed light. The most common method is based on degenerate parametric down conversion using nonlinear crystals that contain second order ($\chi^{(2)}$) susceptibility. When a $\chi^{(2)}$ nonlinear crystal is pumped with photons of frequency $\omega_p$, some of these pump photons get converted into a pair of photons---of frequencies $\omega_p/2$, which are in the single-mode squeezed vacuum state of (\ref{singmosv}).

The connection between squeezing and entanglement is unveiled when two single-mode squeezed vacuum beams are mixed on a beam splitter of the type described in (\ref{eq:UxBS}). The state that results past the beamsplitter is given by the two-mode squeezed vacuum state
\begin{align}
\label{tmsv}
|\xi\rangle&=\hat{S}_2(\xi)|0\rangle_{a_1}|0\rangle_{b_1} \ (\xi=r e^{i (\theta+\pi/2)})\nonumber\\
&=\frac{1}{\cosh r}\sum_{0}^{\infty}(-1)^ne^{in(\theta+\pi/2)}(\tanh r)^n|n\rangle_a|n\rangle_b,
\end{align}
where $\hat{S}_2(\xi)=\exp{\left(\xi \hat{a}_{1}^{\dagger}\hat{b}_1^{\dagger}-\xi^{*}\hat{a}_1\hat{b}_1\right)}$ is the two-mode squeezing operator. This state is mode-entangled as the state of the two modes cannot be written in a separable form. The two-mode squeezing operator can itself also be implemented using a non-degenerate parametric conversion process, where once again a strong pump emitting photons at frequency $\omega_p$ interacts with a nonlinear crystal containing a second order nonlinearity, generating pairs of photons at frequencies $\omega_{a_1}$ and $\omega_{b_1}$, such that $\omega_{a_1}+\omega_{b_1}=\omega_p$.

The scheme of mixing coherent light with squeezed vacuum light, originally considered by Caves, also generates mode entanglement past the first beamsplitter. This provides an alternative explanation for the sub-shot-noise phase precision capabilities of the scheme. We now describe the interferometry with coherent light mixed with squeezed vacuum light in a bit more detail, and show its stronger connection to maximal mode-entanglement of the type present in the $N00N$ states. 

% The very first letter is a 2 line initial drop letter followed% by the rest of the first word in caps.% % form to use if the first word consists of a single letter:% \IEEEPARstart{A}{demo} file is ....% % form to use if you need the single drop letter followed by% normal text (unknown if ever used by IEEE):% \IEEEPARstart{A}{}demo file is ....% % Some journals put the first two words in caps:% \IEEEPARstart{T}{his demo} file is ....% % Here we have the typical use of a "T" for an initial drop letter% and "HIS" in caps to complete the first word.

\subsection{Coherent-mixed with squeezed vacuum light interferometry}
\label{onohofsilb}
\begin{figure}[hbt] 
\begin{center}%Frame position
\includegraphics[scale=0.35]{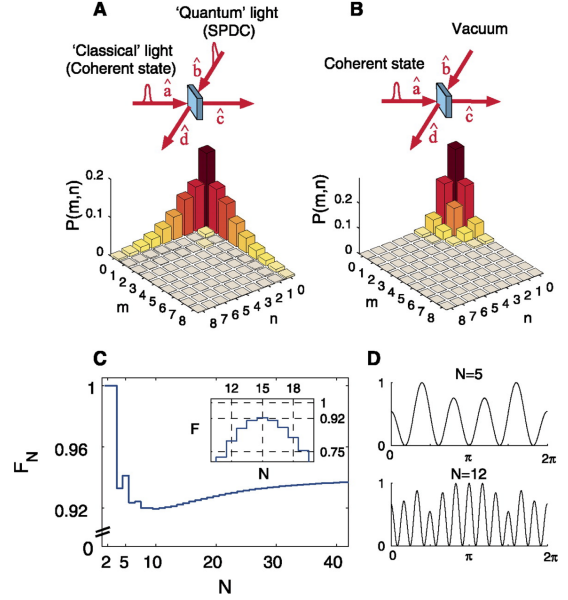}
\end{center}
\caption{Theoretical properties of the states generated in the interferometry with coherent state mixed with squeezed vacuum state. (A) Bar heights represent the probability for m, n photons in output modes c, d, respectively. The corner-like shape illustrates the tendency of all the photons to collectively exit from the same output port, exhibiting Schr{\"o}dinger cat-like behavior. The effect occurs for arbitrary photon fluxes and is demonstrated here using $1.2$ photons per pulse from each input. (B) The same as (A) but using only the coherent state, shown for comparison. The photons at the outputs are clearly in a separable (unentangled) state. (C) Fidelity $F_N$ versus $N$ in an ideal setup. The pair-amplitude ratio $\gamma\equiv |\alpha|^2/r$, which maximizes the $N00N$ state overlap, was chosen separately for each $N$. Optimal fidelity is always larger than 0.92, and it approaches 0.943 asymptotically for large N. The inset shows that when $\gamma$ is optimized for $N = 15$, the fidelity for nearby $N$ is also high. In this case $F > 0.75$ for $N = 12$ to 19, simultaneously. (D) Simulated $N$-fold coincidences as a function of Mach-Zehnder phase for $N = 5$ or 12, demonstrating $N$-fold super-resolution~\cite{AAS10}.}
\label{aas_fig}
\end{figure}

The interferometry based on mixing coherent light and squeezed vacuum light, as mentioned before, was where the possibility of sub-shotnoise phase estimation was originally unearthed. This scheme has been revisited recently. Hofmann and Ono~\cite{HO07} showed that when these inputs are mixed in equal intensities, namely such that $\sinh^2 r=|\alpha|^2=\overline{n}/2$ (for any value of average photon number $\overline{n}$), then the state that results past the mixing splitter is such that each $N$-photon component in the state has a fidelity higher than $90\%$ with the corresponding $N00N$ state. Therefore, this scheme has been widely used to generate $N00N$ states in experiments. Afek et al.~\cite{AAS10, IARAS12} used this scheme to thereby generate $N00N$ states of up to $N=5$; the state of the art in the generation of such states. Fig.~\ref{aas_fig} shows a schematic of this interferometry, photon number distribution of an $N$-photon component in the generated state, its $N00N$ fidelity, and the observed $N$-fold coincidence fringes.

Further, Pezze and Smerzi~\cite{PS08} calculated the classical Cramer-Rao bound for the interferometer with coherent light and squeezed vacuum light along with photon number detection at the output, and found it to be:
\begin{equation}
F_{\rm Cl}=|\alpha|^2e^{2r}+\sinh^2 r.
\end{equation}
When the average photon numbers of the two inputs are about the same, i.e., $\sinh^2 r=|\alpha|^2=\overline{n}/2$, the classical Fisher information is approximately $\overline{n}^2+\overline{n}/2$, which results in Heisenberg scaling for the phase precision, namely $\Delta\varphi=1/(\sqrt{\nu}\overline{n})$, where $\nu$ is the number of data points gathered from measurement identical copies of the state. It has been shown that photon number parity also attains the same phase precision at the Heisenberg scaling in this interferometry~\cite{SALD11}.

\begin{figure*}[hbt] 
\begin{center}%Frame position
\includegraphics[scale=0.85]{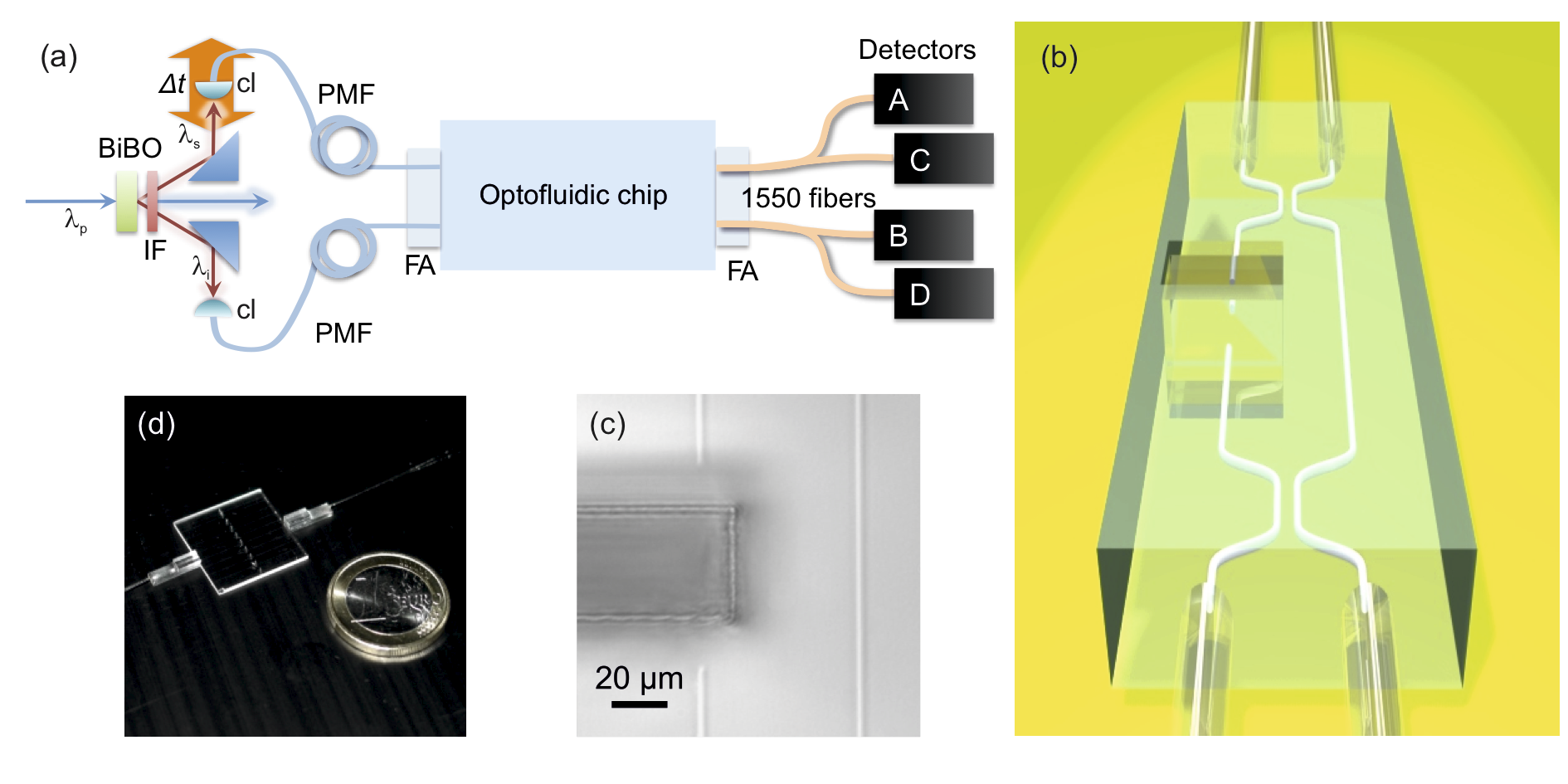}
\end{center}
\caption{Quantum metrology in an optofluidic device. (a) Schematic of the experimental setup: A pump laser at $\lambda_p=392.5$nm generates pairs of downconverted photons at $\lambda_s=\lambda_i=785$nm in a BiBO crystal. IF: interference filter, cl: collection lenses, PMF: polarization maintaining fibers, and FA: fiber array. (b) Schematic of the MZI interfaced to the microchannel. The fluidic channel has rectangular cross-section $500{\rm \mu m}\times 55{\rm \mu m}$ and extends from the top to the bottom surface of the glass substrate ($1{\rm mm}$ thickness). The MZI consists of two 50:50 directional couplers and has two arms of equal geometric length; one waveguide crosses perpendicular to the micro channel, while the other passes externally. (c) Top image of the optical-fluidic interface. (d) Picture of the device with several interferometers and micro channels on chip, together with the fiber arrays for coupling input and output light~\cite{CLMPN12}.}
\label{obrien1}
\end{figure*}

\section{Quantum technologies with entangled photons}
\label{qt}

In this section, we review some recent experiments that have demonstrated the enhanced sensing and imaging capabilities of entangled photons in interferometry with the $N00N$ states. In particular, these experiments focus on the small photon-number regime, which is relevant for sensing and imaging delicate material systems such as biological specimen, single molecules, cold quantum gases and atomic ensembles. We also discuss a recent experiment based on the $N00N$ state for enhanced spatial resolution for applications in quantum lithography.

%that state interferometry for application in sub-shotnoise and
%Heisenberg-limited metrology and sensing, sub-shotnoise imaging and
%sub-Rayleigh imaging.
 
\subsection{Quantum metrology and sensing}
Several experiments based on the $N00N$ states have demonstrated phase estimation beyond the shot-noise limit, and achieving the Heisenberg limit. Here, we briefly mention about two experiments, which have used $N00N$ states to measure useful quantities mapped on to the optical phase under realistic conditions of photon loss and other decoherence. The first one is by Crespi et al.~\cite{CLMPN12} (see Fig.~\ref{obrien1}), where $N=2$ $N00N$ states were used to measure the concentration of a blood protein in an aqueous buffer solution. The experiment used an opto-fluidic device, which consists of a waveguide interferometer whose one arm passes through a microfluidic channel containing the solution. The concentration-dependent refractive index of the solution causes a relative phase shift between the two arms of the interferometer, which is then detected using coincidence photon number detection. The $N=2$ $N00N$ states were generated using Hong-Ou-Mandel interference~\cite{GK05} with entangled photon pairs from a parametric down conversion source. At the output, an array of telecommunication optical fibers were used to collect the photons, which were then detected with coincidence detection using four single-photon avalanche photo-diodes. The photons were detected with a fringe visibility of about $87\%$ in the case were the micro channel had a transmissivity of only about $61\%$ due to photon loss. The experiment achieved a sensitivity below the shot-noise limit.

\begin{figure}[hbt] 
\begin{center}%Frame position
\includegraphics[scale=0.35]{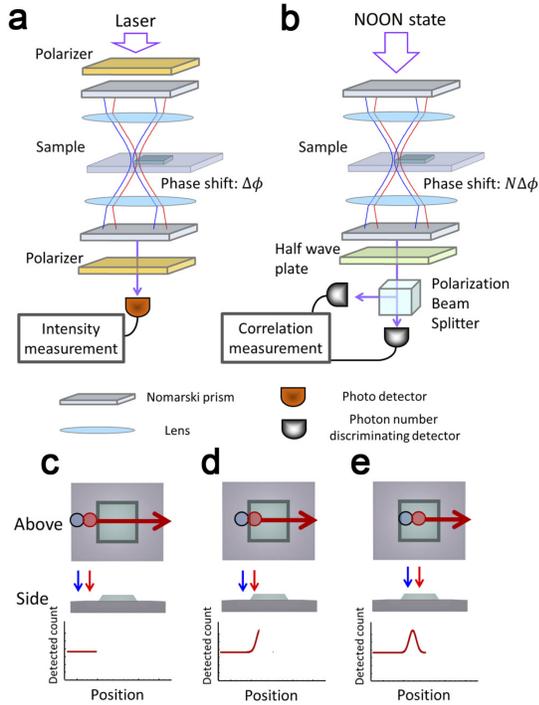}
\end{center}
\caption{Illustration of {\bf (a)} laser-confocal-type differential interference contrast microscope, {\bf (b)} the entanglement-enhanced microscope. The red and blue lines indicate horizontally and vertically polarized light. {\bf (c), (d)} and {\bf (e)} The change in the signal while the sample is scanned~\cite{OOT13}.
%Construction process of the signal by scanning the sample. 
}
\label{ono1}
\end{figure}

\begin{figure}[hbt] 
\begin{center}%Frame position
\includegraphics[scale=0.35]{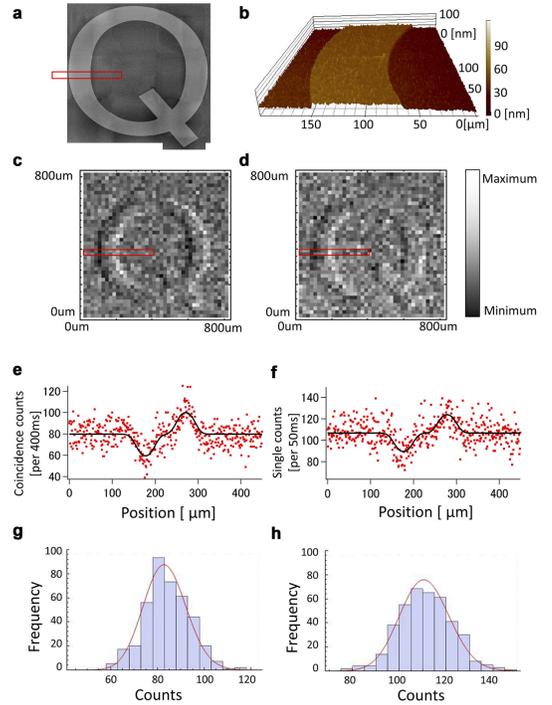}
\end{center}
\caption{(a) Atomic force microscope (AFM) image of a glass plate sample (BK7) on whose surface a `Q' shape is carved in relief with an ultra-thin step using optical lithography. (b) The section of the AFM image of the sample, which is the area outlined in red in (a). The height of the step is estimated to be 17.3 nm from this data. (c) The image of the sample using an entanglement-enhanced microscope where two photon entangled state is used to illuminate the sample. (d) The image of the sample using single photons (a classical light source). (e) and (f) are 1D fine scan data for the area outlined in red in (c) and (d) for the same photon number of 920. The measurement was made at a bias phase of 0.41 (e) and 0.66 (f), where optimal bias phase are 0.38 and 0.67, respectively~\cite{OOT13}.}
\label{ono3}
\end{figure}

In another experiment, Wolfgramm et al.~\cite{WVBGN13} used $N=2$ polarization $N00N$ states and Faraday rotation to probe a Rubidium atomic spin ensemble in a non-destructive manner. Atomic spins ensembles find application in optical quantum memory, quantum-enhanced atom interferometry, etc. Such atomic spin ensembles, when interacted with via optical measurements, e.g., to store or readout quantum information in a quantum memory or to produce spin-squeezing in atom interferometry, inherently suffer from scattering induced depolarization noise. Also, there is photon loss due to the scattering of the optical probes off the ensemble. In order to minimize loss, the experiment generated narrowband $N00N$ states of about $90\%$ fidelity and purity, at a frequency detuned four Doppler widths from the nearest Rb-85 resonance containing matter-resonant indistinguishable photons. The photons at the output were detected using condense photon number detection with a fringe visibility of $>90\%$. The experiment achieved a sensitivity that was five standard deviations better than the shot-noise limit.

\subsection{Quantum imaging}
Another important application of optical phase measurement is that of microscopy and imaging. In biology, the technique of differential interference contrast microscopy is widely used to image biological samples. The depth resolution of the images produced by this technique is related to the signal-to-noise ratio (SNR) of the measurement. In the case of classical laser-light-based imaging, for a given light intensity, this is limited by the shot-noise limit in phase precision. While one way to enhance the SNR is to raise the illumination power, this might have undesirable effects on delicate, photosensitive samples such as biological tissues, ice crystals, etc. Quantum metrology, however, can provide an enhancement to the SNR without having to increase the illumination power, and therefore could be of significant help in this scenario. In one of the first works on the use of quantum metrology for phase imaging, Brida et al.~\cite{BGB10} showed that entangled photon pairs can provide sub-shot-noise imaging of absorbing samples. Later, Taylor et al.~\cite{TJDK13} showed that squeezed light could be used to achieve sub-shot-noise sensitivities in micro-particle tracking, with applications in tracking diffusive biological specimen in realtime. 

We describe two recent experiments that have used entangled photons for phase super-sensitive imaging. The first one is by Ono et al.~\cite{OOT13}, where $N=2$ $N00N$ states were used in a laser confocal microscope in conjunction with a differential interference contrast microscope (LCM-DIM) to demonstrate quantum-enhanced microscopy. Figures~\ref{ono1} and~\ref{ono3} show their experimental setup and results, respectively. The LCM-DIM works based on polarization interferometry, where the $H$ and $V$ modes are separated using a polarization beamsplitter or a Nomarski prism, and made to pass through different spatial parts of the sample. These modes, depending on the local refractive index and the structure of the sample, experience different phase shifts, whose difference is then measured at the output. The experiment used $N=2$ polarization $N00N$ states generated via Hong-Ou-Mandel interference with about 98\% fidelity, and the photons were detected at the output using photon number parity measurement with a fringe visibility of about $95\%$. The microscope attained an SNR $1.35$-times better than the shot noise limit.

In another experiment, Israel et al.~\cite{IRS14} used $N=2 $ and $N=3 $ $N00N$ states in quantum polarization light microscopy (QPLM) to image a quartz crystal. In QPLM, a birefringent sample causes the $H$ and $V$ modes to experience a differential phase shift, which is then measured at the output to image the sample. The $N00N$ states for the experiment were generated from the mixing of coherent light and squeezed vacuum light in equal intensities discussed in Section~\ref{onohofsilb}, and the photons were detected using an array of single photon counting modules. The experiment achieved quantum-enhanced imaging with sensitivities close to the Heisenberg limit.

\subsection{Quantum lithography}

Lithography relies on the creation and detection of spatial interference fringes to etch ultra fine features on a chip. While classical light lithography is limited by the Rayleigh diffraction limit, as mentioned before, the $N00N$ states can beat this limit---a result known as super-resolution~\cite{KBD04,KBAWBD01, BKABCD00}. A few independent experiments with $N=2$ $N00N$ states had earlier demonstrated this result, however, it was realized that the $N00N$ state lithography suffers from the problem that the efficiency of detection $N$ photons in the same spatial location decreases exponentially in $N$. 

In a new theoretical development, a counter-measure was suggested based on the optical centroid measurement~\cite{Tsa09}, which does not require all the photons to arrive at the same spatial point. The optical centroid measurement is based on an array of detectors that keep track of every $N$ photon detection event irrespective of which detectors fired. Then the average position of the photons is obtained via post-processing. In a recent experiment, using $N=2,\ 3,\ 4$ $N00N$ states and the optical centroid measurement, Rozema et al.~\cite{RBMOF13} for the first time demonstrated a scalable implementation of quantum super-resolving interferometry with a visibility of interference fringes nearly independent of $N$.

\section{Discussion}
\label{disc}

In this paper, we presented a brief overview of quantum optical metrology, with an emphasis on quantum technologies that have been demonstrated with the $N00N$ states. We introduced some basic concepts of quantum interferometry and quantum parameter estimation, and the notions of entanglement and squeezing.  We then presented an interferometric scheme based on the mixing of coherent and squeezed vacuum light on a beam splitter, which generates an output whose $N$-photon components are approximately $N00N$ states. We then presented some state-of-the-art experiments that use $N00N$ states generated using this technique (or other techniques in some cases) for technological applications such as quantum-enhanced biosensing, imaging, and spatial resolution lithography.

This review is by no means representative of everything there is to quantum optical metrology. For a comprehensive review of the field, please refer~\cite{DJK14, Kol14}. We did not discuss the approach to quantum metrology via the Bayesian method~\cite{Hol82}, where the unknown parameter is assumed to be inherently random, and thus distributed according to an unknown probability distribution. Optimal states in this paradigm have also been identified, and adaptive protocols have been designed, which implement the optimal measurements for such states based on measurement settings that continually changed based on the results previously obtained~\cite{XHBWP11, HBBWP07, BW00, Wise95}. Also, we did not go into the details of how the effects of photon loss and decoherence such as collective dephasing noise due to the thermal motion of optical components or laser noise, etc, are handled in optical metrology. A large body of work in the recent literature has dealt with identifying useful lower bounds on phase precision in the presence of such decoherence~\cite{DKM12, EFD11}. Further, optimal quantum states of light that attain these bounds in the presence of decoherence have been identified in the asymptotic limit of a large number of photons~\cite{KCD14}.

It must be mentioned here that the $N00N$ states are highly susceptible to photon loss, or other types of decoherence in the limit of a large photon number N. Nevertheless, the use of the $N00N$ state in the experiments discussed here is justified, since the $N00N$ states still remain optimal for relatively small photon numbers. In fact, in noisy, decoherence-ridden interferometry, given a large finite photon number constraint, it has been recently shown that the best strategy for phase estimation is to divide the total number of photons into smaller independent packets or "clusters", where each cluster is prepared in a $N00N$ state~\cite{KCD14}. These clusters are then to be sent through the interferometer one at a time. (The optimal size of the clusters will depend on the decoherence strength inside the interferometer.) In addition, alternatives to the $N00N$ states, of the form $(|M\rangle|N-M\rangle+|N-M\rangle|M\rangle)/2$, where $M\leq N$ (also known as the $mm'$ states) have been proposed for interferometry in the presence of photon loss~\cite{JBWKLD12, HWD08}. Such states, when suitably chosen, offer the same benefits as the $N00N$ states, while being more robust against photon loss than the latter. Both the $N00N$ states and the $mm'$ states of a moderate number of photons, have been shown to perform optimally in the presence of collective dephasing noise~\cite{RJD13}. Therefore, the $mm'$ states may provide a way to perform quantum metrology in the presence of both photon loss and collective dephasing noise in suitable regimes of photon numbers and the decoherence parameters.

% if have a single appendix:%\appendix[Proof of the Zonklar Equations]% or%\appendix  % for no appendix heading% do not use \section anymore after \appendix, only \section*% is possibly needed

% use appendices with more than one appendix% then use \section to start each appendix% you must declare a \section before using any% \subsection or using \label (\appendices by itself% starts a section numbered zero.)

\appendices{ }

% you can choose not to have a title for an appendix% if you want by leaving the argument blank


\section*{Acknowledgment}

KPS would like to thank the Graduate School of Louisiana State University for the 2014-2015 Dissertation Year Fellowship. JPD would like to thank the Air Force Office of Scientific Research, the Army Research Office, and the National Science Foundation.

% Can use something like this to put references on a page% by themselves when using endfloat and the captionsoff option.\ifCLASSOPTIONcaptionsoff\newpage{}

% trigger a \newpage just before the given reference% number - used to balance the columns on the last page% adjust value as needed - may need to be readjusted if% the document is modified later%\IEEEtriggeratref{8}% The "triggered" command can be changed if desired:%\IEEEtriggercmd{\enlargethispage{-5in}}

% references section \bibliographystyle{plain}
%\bibliography{Ref-1}
\bibliographystyle{ieeetran}
\bibliography{Ref-1.bib}

\begin{thebibliography}{10}
\providecommand{\url}[1]{#1}
\csname url@samestyle\endcsname
\providecommand{\newblock}{\relax}
\providecommand{\bibinfo}[2]{#2}
\providecommand{\BIBentrySTDinterwordspacing}{\spaceskip=0pt\relax}
\providecommand{\BIBentryALTinterwordstretchfactor}{4}
\providecommand{\BIBentryALTinterwordspacing}{\spaceskip=\fontdimen2\font plus
\BIBentryALTinterwordstretchfactor\fontdimen3\font minus
  \fontdimen4\font\relax}
\providecommand{\BIBforeignlanguage}[2]{{%
\expandafter\ifx\csname l@#1\endcsname\relax
\typeout{** WARNING: IEEEtran.bst: No hyphenation pattern has been}%
\typeout{** loaded for the language `#1'. Using the pattern for}%
\typeout{** the default language instead.}%
\else
\language=\csname l@#1\endcsname
\fi
#2}}
\providecommand{\BIBdecl}{\relax}
\BIBdecl

\bibitem{Shor97}
P.~W. Shor, ``Polynomial-time algorithms for prime factorization and discrete
  logarithms on a quantum computer,'' \emph{SIAM J. Sci. Statist. Comput.},
  vol.~26, p. 1484, 1997.

\bibitem{Grov96}
L.~K. Grover, ``A fast quantum mechanical algorithm for database search,'' in
  \emph{Proceedings, 28th Annual ACM Symposium on the Theory of
  Computing}.\hskip 1em plus 0.5em minus 0.4em\relax ACM, 1996, pp. 212--219.

\bibitem{BBCJPW93}
\BIBentryALTinterwordspacing
C.~H. Bennett, G.~Brassard, C.~Cr\'epeau, R.~Jozsa, A.~Peres, and W.~K.
  Wootters, ``Teleporting an unknown quantum state via dual classical and
  einstein-podolsky-rosen channels,'' \emph{Phys. Rev. Lett.}, vol.~70, pp.
  1895--1899, Mar 1993. [Online]. Available:
  \url{http://link.aps.org/doi/10.1103/PhysRevLett.70.1895}
\BIBentrySTDinterwordspacing

\bibitem{Eke91}
\BIBentryALTinterwordspacing
A.~K. Ekert, ``Quantum cryptography based on bell's theorem,'' \emph{Phys. Rev.
  Lett.}, vol.~67, pp. 661--663, Aug 1991. [Online]. Available:
  \url{http://link.aps.org/doi/10.1103/PhysRevLett.67.661}
\BIBentrySTDinterwordspacing

\bibitem{BB84}
C.~H. Bennett and G.~Brassard, ``Quantum cryptography: Public key distribution
  and coin tossing,'' in \emph{Proceedings of IEEE International Conference on
  Computers, Systems and Signal Processing}.\hskip 1em plus 0.5em minus
  0.4em\relax IEEE, 1984, pp. pp. 175--179.

\bibitem{GLM06}
\BIBentryALTinterwordspacing
V.~Giovannetti, S.~Lloyd, and L.~Maccone, ``Quantum metrology,'' \emph{Phys.
  Rev. Lett.}, vol.~96, p. 010401, Jan 2006. [Online]. Available:
  \url{http://link.aps.org/doi/10.1103/PhysRevLett.96.010401}
\BIBentrySTDinterwordspacing

\bibitem{PRRH11}
\BIBentryALTinterwordspacing
M.~Pitkin, S.~Reid, S.~Rowan, and J.~Hough, ``Gravitational wave detection by
  interferometry (ground and space),'' \emph{Living Reviews in Relativity},
  vol.~14, no.~5, 2011. [Online]. Available:
  \url{http://www.livingreviews.org/lrr-2011-5}
\BIBentrySTDinterwordspacing

\bibitem{GLM01}
\BIBentryALTinterwordspacing
V.~Giovannetti, S.~Lloyd, and L.~Maccone, ``{Quantum-enhanced positioning and
  clock synchronization},'' \emph{Nature}, vol. 412, no. 6845, pp. 417--419,
  Jul. 2001. [Online]. Available: \url{http://dx.doi.org/10.1038/35086525}
\BIBentrySTDinterwordspacing

\bibitem{HMPEP97}
\BIBentryALTinterwordspacing
S.~F. Huelga, C.~Macchiavello, T.~Pellizzari, A.~K. Ekert, M.~B. Plenio, and
  J.~I. Cirac, ``Improvement of frequency standards with quantum
  entanglement,'' \emph{Phys. Rev. Lett.}, vol.~79, pp. 3865--3868, Nov 1997.
  [Online]. Available:
  \url{http://link.aps.org/doi/10.1103/PhysRevLett.79.3865}
\BIBentrySTDinterwordspacing

\bibitem{WVBGN13}
\BIBentryALTinterwordspacing
F.~Wolfgramm, C.~Vitelli, F.~A. Beduini, N.~Godbout, and M.~W. Mitchell,
  ``{Entanglement-enhanced probing of a delicate material system},'' \emph{Nat
  Photon}, vol.~7, no.~1, pp. 28--32, Jan. 2013. [Online]. Available:
  \url{http://dx.doi.org/10.1038/nphoton.2012.300
  http://www.nature.com/nphoton/journal/v7/n1/abs/nphoton.2012.300.html\#supplementary-information}
\BIBentrySTDinterwordspacing

\bibitem{CLMPN12}
\BIBentryALTinterwordspacing
A.~Crespi, M.~Lobino, J.~C.~F. Matthews, A.~Politi, C.~R. Neal, R.~Ramponi,
  R.~Osellame, and J.~L. O'Brien, ``Measuring protein concentration with
  entangled photons,'' \emph{Applied Physics Letters}, vol. 100, no.~23, p.
  233704, 2012. [Online]. Available:
  \url{http://scitation.aip.org/content/aip/journal/apl/100/23/10.1063/1.4724105}
\BIBentrySTDinterwordspacing

\bibitem{JLGD13}
\BIBentryALTinterwordspacing
K.~Jiang, H.~Lee, C.~C. Gerry, and J.~P. Dowling, ``Super-resolving quantum
  radar: Coherent-state sources with homodyne detection suffice to beat the
  diffraction limit,'' \emph{Journal of Applied Physics}, vol. 114, no.~19, p.
  193102, 2013. [Online]. Available:
  \url{http://scitation.aip.org/content/aip/journal/jap/114/19/10.1063/1.4829016}
\BIBentrySTDinterwordspacing

\bibitem{GAWLLD10}
Y.~Gao, P.~M. Anisimov, C.~F. Wildfeuer, J.~Luine, H.~Lee, and J.~P. Dowling,
  ``Super-resolution at the shot-noise limit with coherent states and
  photon-number-resolving detectors,'' \emph{J. Opt. Soc. Am. B}, vol.~27,
  no.~6, pp. A170--A174, Jun 2010.

\bibitem{Sh07}
Y.~Shih, ``Quantum imaging,'' \emph{Selected Topics in Quantum Electronics,
  IEEE Journal of}, vol.~13, no.~4, pp. 1016--1030, July 2007.

\bibitem{LGB02}
\BIBentryALTinterwordspacing
L.~A. Lugiato, A.~Gatti, and E.~Brambilla, ``Quantum imaging,'' \emph{Journal
  of Optics B: Quantum and Semiclassical Optics}, vol.~4, no.~3, p. S176, 2002.
  [Online]. Available: \url{http://stacks.iop.org/1464-4266/4/i=3/a=372}
\BIBentrySTDinterwordspacing

\bibitem{KBD04}
\BIBentryALTinterwordspacing
P.~Kok, S.~L. Braunstein, and J.~P. Dowling, ``Quantum lithography,
  entanglement and heisenberg-limited parameter estimation,'' \emph{Journal of
  Optics B: Quantum and Semiclassical Optics}, vol.~6, no.~8, p. S811, 2004.
  [Online]. Available: \url{http://stacks.iop.org/1464-4266/6/i=8/a=029}
\BIBentrySTDinterwordspacing

\bibitem{KBAWBD01}
\BIBentryALTinterwordspacing
P.~Kok, A.~N. Boto, D.~S. Abrams, C.~P. Williams, S.~L. Braunstein, and J.~P.
  Dowling, ``Quantum-interferometric optical lithography: Towards arbitrary
  two-dimensional patterns,'' \emph{Phys. Rev. A}, vol.~63, p. 063407, May
  2001. [Online]. Available:
  \url{http://link.aps.org/doi/10.1103/PhysRevA.63.063407}
\BIBentrySTDinterwordspacing

\bibitem{BKABCD00}
\BIBentryALTinterwordspacing
A.~N. Boto, P.~Kok, D.~S. Abrams, S.~L. Braunstein, C.~P. Williams, and J.~P.
  Dowling, ``Quantum interferometric optical lithography: Exploiting
  entanglement to beat the diffraction limit,'' \emph{Phys. Rev. Lett.},
  vol.~85, pp. 2733--2736, Sep 2000. [Online]. Available:
  \url{http://link.aps.org/doi/10.1103/PhysRevLett.85.2733}
\BIBentrySTDinterwordspacing

\bibitem{He76}
C.~W. Helstrom, ``Quantum detection and estimation theory,'' \emph{Academic
  Press}, 1976.

\bibitem{BC94}
\BIBentryALTinterwordspacing
S.~L. Braunstein and C.~M. Caves, ``Statistical distance and the geometry of
  quantum states,'' \emph{Phys. Rev. Lett.}, vol.~72, pp. 3439--3443, May 1994.
  [Online]. Available:
  \url{http://link.aps.org/doi/10.1103/PhysRevLett.72.3439}
\BIBentrySTDinterwordspacing

\bibitem{BCM96}
S.~L. Braunstein, C.~M. Caves, and G.~J. Milburn, ``Generalized uncertainty
  relations: Theory, examples, and lorentz invariance,'' \emph{Ann. Phys.},
  vol. 247, p. 135, 1996.

\bibitem{HB93}
\BIBentryALTinterwordspacing
M.~J. Holland and K.~Burnett, ``Interferometric detection of optical phase
  shifts at the heisenberg limit,'' \emph{Phys. Rev. Lett.}, vol.~71, pp.
  1355--1358, Aug 1993. [Online]. Available:
  \url{http://link.aps.org/doi/10.1103/PhysRevLett.71.1355}
\BIBentrySTDinterwordspacing

\bibitem{Ca81}
\BIBentryALTinterwordspacing
C.~M. Caves, ``Quantum-mechanical noise in an interferometer,'' \emph{Phys.
  Rev. D}, vol.~23, pp. 1693--1708, Apr 1981. [Online]. Available:
  \url{http://link.aps.org/doi/10.1103/PhysRevD.23.1693}
\BIBentrySTDinterwordspacing

\bibitem{BS84}
\BIBentryALTinterwordspacing
R.~S. Bondurant and J.~H. Shapiro, ``Squeezed states in phase-sensing
  interferometers,'' \emph{Phys. Rev. D}, vol.~30, pp. 2548--2556, Dec 1984.
  [Online]. Available: \url{http://link.aps.org/doi/10.1103/PhysRevD.30.2548}
\BIBentrySTDinterwordspacing

\bibitem{Do08}
J.~P. Dowling, ``Quantum optical metrology---the lowdown on high-{N}00{N}
  states,'' \emph{Contemp. Phys.}, vol.~49, no.~2, pp. 125--143, 2008.

\bibitem{BW00}
\BIBentryALTinterwordspacing
D.~W. Berry and H.~M. Wiseman, ``Optimal states and almost optimal adaptive
  measurements for quantum interferometry,'' \emph{Phys. Rev. Lett.}, vol.~85,
  pp. 5098--5101, Dec 2000. [Online]. Available:
  \url{http://link.aps.org/doi/10.1103/PhysRevLett.85.5098}
\BIBentrySTDinterwordspacing

\bibitem{VDGJK14}
\BIBentryALTinterwordspacing
M.~D. Vidrighin, G.~Donati, M.~G. Genoni, X.-M. Jin, W.~S. Kolthammer, M.~S.
  Kim, A.~Datta, M.~Barbieri, and I.~A. Walmsley, ``{Joint estimation of phase
  and phase diffusion for quantum metrology},'' \emph{Nat Commun}, vol.~5, Apr.
  2014. [Online]. Available: \url{http://dx.doi.org/10.1038/ncomms4532
  10.1038/ncomms4532}
\BIBentrySTDinterwordspacing

\bibitem{XHBWP11}
\BIBentryALTinterwordspacing
G.~Y. Xiang, B.~L. Higgins, D.~W. Berry, H.~M. Wiseman, and G.~J. Pryde,
  ``Entanglement-enhanced measurement of a completely unknown optical phase,''
  \emph{Nat Photon}, vol.~5, pp. 43--47, Jan 2011. [Online]. Available:
  \url{http://dx.doi.org/10.1038/nphoton.2010.268}
\BIBentrySTDinterwordspacing

\bibitem{AAS10}
\BIBentryALTinterwordspacing
I.~Afek, O.~Ambar, and Y.~Silberberg, ``High-{N}00{N} states by mixing quantum
  and classical light,'' \emph{Science}, vol. 328, no. 5980, pp. 879--881,
  2010. [Online]. Available:
  \url{http://www.sciencemag.org/content/328/5980/879.abstract}
\BIBentrySTDinterwordspacing

\bibitem{HBBWP07}
\BIBentryALTinterwordspacing
B.~L. Higgins, D.~W. Berry, S.~D. Bartlett, H.~M. Wiseman, and G.~J. Pryde,
  ``Entanglement-free {H}eisenberg-limited phase estimation,'' \emph{Nature},
  vol. 450, no. 7168, pp. 393--396, 11 2007. [Online]. Available:
  \url{http://dx.doi.org/10.1038/nature06257}
\BIBentrySTDinterwordspacing

\bibitem{NOOST07}
\BIBentryALTinterwordspacing
T.~Nagata, R.~Okamoto, J.~L. O'Brien, K.~Sasaki, and S.~Takeuchi, ``Beating the
  standard quantum limit with four-entangled photons,'' \emph{Science}, vol.
  316, no. 5825, pp. 726--729, 2007. [Online]. Available:
  \url{http://www.sciencemag.org/content/316/5825/726.abstract}
\BIBentrySTDinterwordspacing

\bibitem{Bo04}
\BIBentryALTinterwordspacing
D.~Bouwmeester, ``Quantum physics: High noon for photons,'' \emph{Nature}, vol.
  429, pp. 139--141, May 2004. [Online]. Available:
  \url{http://dx.doi.org/10.1038/429139a}
\BIBentrySTDinterwordspacing

\bibitem{YC83}
\BIBentryALTinterwordspacing
H.~P. Yuen and V.~W.~S. Chan, ``Noise in homodyne and heterodyne detection,''
  \emph{Opt. Lett.}, vol.~8, no.~3, pp. 177--179, Mar 1983. [Online].
  Available: \url{http://ol.osa.org/abstract.cfm?URI=ol-8-3-177}
\BIBentrySTDinterwordspacing

\bibitem{PB88}
\BIBentryALTinterwordspacing
D.~T. Pegg and S.~M. Barnett, ``Unitary phase operator in quantum mechanics,''
  \emph{EPL (Europhysics Letters)}, vol.~6, no.~6, p. 483, 1988. [Online].
  Available: \url{http://stacks.iop.org/0295-5075/6/i=6/a=002}
\BIBentrySTDinterwordspacing

\bibitem{Wise95}
\BIBentryALTinterwordspacing
H.~M. Wiseman, ``Adaptive phase measurements of optical modes: Going beyond the
  marginal {$Q$} distribution,'' \emph{Phys. Rev. Lett.}, vol.~75, pp.
  4587--4590, Dec 1995. [Online]. Available:
  \url{http://link.aps.org/doi/10.1103/PhysRevLett.75.4587}
\BIBentrySTDinterwordspacing

\bibitem{RLMN05}
\BIBentryALTinterwordspacing
D.~Rosenberg, A.~E. Lita, A.~J. Miller, and S.~W. Nam, ``Noise-free
  high-efficiency photon-number-resolving detectors,'' \emph{Phys. Rev. A},
  vol.~71, p. 061803, Jun 2005, arXiv:quant-ph/0506175v1. [Online]. Available:
  \url{http://link.aps.org/doi/10.1103/PhysRevA.71.061803}
\BIBentrySTDinterwordspacing

\bibitem{KYS08}
\BIBentryALTinterwordspacing
B.~E. Kardyna, Z.~L. Yuan, and A.~J. Shields, ``An avalanche-photodiode-based
  photon-number-resolving detector,'' \emph{Nature Photonics}, vol.~2, pp.
  425--428, Jun 2008. [Online]. Available:
  \url{http://dx.doi.org/10.1038/nphoton.2008.101}
\BIBentrySTDinterwordspacing

\bibitem{GM10}
C.~C. Gerry and J.~Mimih, ``The parity operator in quantum optical metrology,''
  \emph{Contemp. Phys.}, vol.~51, no.~6, pp. 497--511, 2010.

\bibitem{KCD14}
S.~I. Knysh, E.~H. Chen, and G.~A. Durkin, ``{T}rue {L}imits to {P}recision via
  {U}nique {Q}uantum {P}robe,'' 2014, arXiv:1402.0495v1.

\bibitem{DKM12}
\BIBentryALTinterwordspacing
R.~Demkowicz-Dobrzanski, J.~Kolodynski, and M.~Guta, ``{The elusive Heisenberg
  limit in quantum-enhanced metrology},'' \emph{Nat Commun}, vol.~3, p. 1063,
  Sep. 2012. [Online]. Available: \url{http://dx.doi.org/10.1038/ncomms2067
  http://www.nature.com/ncomms/journal/v3/n9/suppinfo/ncomms2067\_S1.html}
\BIBentrySTDinterwordspacing

\bibitem{KSD11}
\BIBentryALTinterwordspacing
S.~Knysh, V.~N. Smelyanskiy, and G.~A. Durkin, ``Scaling laws for precision in
  quantum interferometry and the bifurcation landscape of the optimal state,''
  \emph{Phys. Rev. A}, vol.~83, p. 021804, Feb 2011. [Online]. Available:
  \url{http://link.aps.org/doi/10.1103/PhysRevA.83.021804}
\BIBentrySTDinterwordspacing

\bibitem{EFD11}
\BIBentryALTinterwordspacing
B.~M. Escher, R.~L. {de Matos Filho}, and L.~Davidovich, ``{General framework
  for estimating the ultimate precision limit in noisy quantum-enhanced
  metrology},'' \emph{Nat Phys}, vol.~7, no.~5, pp. 406--411, May 2011.
  [Online]. Available: \url{http://dx.doi.org/10.1038/nphys1958
  http://www.nature.com/nphys/journal/v7/n5/abs/nphys1958.html\#supplementary-information}
\BIBentrySTDinterwordspacing

\bibitem{LHLKMM09}
\BIBentryALTinterwordspacing
T.-W. Lee, S.~D. Huver, H.~Lee, L.~Kaplan, S.~B. McCracken, C.~Min, D.~B.
  Uskov, C.~F. Wildfeuer, G.~Veronis, and J.~P. Dowling, ``Optimization of
  quantum interferometric metrological sensors in the presence of photon
  loss,'' \emph{Phys. Rev. A}, vol.~80, p. 063803, Dec 2009. [Online].
  Available: \url{http://link.aps.org/doi/10.1103/PhysRevA.80.063803}
\BIBentrySTDinterwordspacing

\bibitem{DBSLWBW09}
\BIBentryALTinterwordspacing
U.~Dorner, R.~Demkowicz-Dobrzanski, B.~J. Smith, J.~S. Lundeen, W.~Wasilewski,
  K.~Banaszek, and I.~A. Walmsley, ``Optimal quantum phase estimation,''
  \emph{Phys. Rev. Lett.}, vol. 102, p. 040403, Jan 2009. [Online]. Available:
  \url{http://link.aps.org/doi/10.1103/PhysRevLett.102.040403}
\BIBentrySTDinterwordspacing

\bibitem{IRS14}
\BIBentryALTinterwordspacing
Y.~Israel, S.~Rosen, and Y.~Silberberg, ``Supersensitive polarization
  microscopy using noon states of light,'' \emph{Phys. Rev. Lett.}, vol. 112,
  p. 103604, Mar 2014. [Online]. Available:
  \url{http://link.aps.org/doi/10.1103/PhysRevLett.112.103604}
\BIBentrySTDinterwordspacing

\bibitem{RBMOF13}
\BIBentryALTinterwordspacing
L.~A. Rozema, J.~D. Bateman, D.~H. Mahler, R.~Okamoto, A.~Feizpour, A.~Hayat,
  and A.~M. Steinberg, ``Scalable spatial superresolution using entangled
  photons,'' \emph{Phys. Rev. Lett.}, vol. 112, p. 223602, Jun 2014,
  arXiv:1312.2012v1. [Online]. Available:
  \url{http://link.aps.org/doi/10.1103/PhysRevLett.112.223602}
\BIBentrySTDinterwordspacing

\bibitem{OOT13}
\BIBentryALTinterwordspacing
T.~Ono, R.~Okamoto, and S.~Takeuchi, ``{An entanglement-enhanced microscope},''
  \emph{Nat Commun}, vol.~4, Sep. 2013. [Online]. Available:
  \url{http://dx.doi.org/10.1038/ncomms3426 10.1038/ncomms3426}
\BIBentrySTDinterwordspacing

\bibitem{WPGNRSL12}
\BIBentryALTinterwordspacing
C.~Weedbrook, S.~Pirandola, R.~Garcia-Patr\'on, N.~J. Cerf, T.~C. Ralph, J.~H.
  Shapiro, and S.~Lloyd, ``Gaussian quantum information,'' \emph{Rev. Mod.
  Phys.}, vol.~84, pp. 621--669, May 2012. [Online]. Available:
  \url{http://link.aps.org/doi/10.1103/RevModPhys.84.621}
\BIBentrySTDinterwordspacing

\bibitem{GK05}
C.~C. Gerry and P.~L. Knight, \emph{Introductory Quantum Optics}.\hskip 1em
  plus 0.5em minus 0.4em\relax Cambridge University Press, 2005.

\bibitem{YMK86}
\BIBentryALTinterwordspacing
B.~Yurke, S.~L. McCall, and J.~R. Klauder, ``Su(2) and su(1,1)
  interferometers,'' \emph{Phys. Rev. A}, vol.~33, pp. 4033--4054, Jun 1986.
  [Online]. Available: \url{http://link.aps.org/doi/10.1103/PhysRevA.33.4033}
\BIBentrySTDinterwordspacing

\bibitem{sakurai}
J.~J. Sakurai, \emph{Modern Quantum Mechanics}.\hskip 1em plus 0.5em minus
  0.4em\relax Addison Wesley, 1994.

\bibitem{SKDL13}
\BIBentryALTinterwordspacing
K.~P. Seshadreesan, S.~Kim, J.~P. Dowling, and H.~Lee, ``Phase estimation at
  the quantum cram\'er-rao bound via parity detection,'' \emph{Phys. Rev. A},
  vol.~87, p. 043833, Apr 2013. [Online]. Available:
  \url{http://link.aps.org/doi/10.1103/PhysRevA.87.043833}
\BIBentrySTDinterwordspacing

\bibitem{SALD11}
K.~P. Seshadreesan, P.~M. Anisimov, H.~Lee, and J.~P. Dowling, ``Parity
  detection achieves the heisenberg limit in interferometry with coherent mixed
  with squeezed vacuum light,'' \emph{New J. Phys.}, vol.~13, no.~8, p. 083026,
  2011.

\bibitem{ARCPHLD10}
\BIBentryALTinterwordspacing
P.~M. Anisimov, G.~M. Raterman, A.~Chiruvelli, W.~N. Plick, S.~D. Huver,
  H.~Lee, and J.~P. Dowling, ``Quantum metrology with two-mode squeezed vacuum:
  Parity detection beats the {H}eisenberg limit,'' \emph{Phys. Rev. Lett.},
  vol. 104, p. 103602, Mar 2010. [Online]. Available:
  \url{http://link.aps.org/doi/10.1103/PhysRevLett.104.103602}
\BIBentrySTDinterwordspacing

\bibitem{Eng96}
\BIBentryALTinterwordspacing
B.-G. Englert, ``Fringe visibility and which-way information: An inequality,''
  \emph{Phys. Rev. Lett.}, vol.~77, pp. 2154--2157, Sep 1996. [Online].
  Available: \url{http://link.aps.org/doi/10.1103/PhysRevLett.77.2154}
\BIBentrySTDinterwordspacing

\bibitem{HHHH09}
\BIBentryALTinterwordspacing
R.~Horodecki, P.~Horodecki, M.~Horodecki, and K.~Horodecki, ``{Q}uantum
  entanglement,'' \emph{Reviews of Modern Physics}, vol.~81, no.~2, pp.
  865--942, June 2009, arXiv:quant-ph/0702225v2. [Online]. Available:
  \url{http://link.aps.org/doi/10.1103/RevModPhys.81.865}
\BIBentrySTDinterwordspacing

\bibitem{BCPSW14}
\BIBentryALTinterwordspacing
N.~Brunner, D.~Cavalcanti, S.~Pironio, V.~Scarani, and S.~Wehner, ``Bell
  nonlocality,'' \emph{Rev. Mod. Phys.}, vol.~86, pp. 419--478, Apr 2014.
  [Online]. Available: \url{http://link.aps.org/doi/10.1103/RevModPhys.86.419}
\BIBentrySTDinterwordspacing

\bibitem{PS09}
\BIBentryALTinterwordspacing
L.~Pezz\'e and A.~Smerzi, ``Entanglement, nonlinear dynamics, and the
  heisenberg limit,'' \emph{Phys. Rev. Lett.}, vol. 102, p. 100401, Mar 2009.
  [Online]. Available:
  \url{http://link.aps.org/doi/10.1103/PhysRevLett.102.100401}
\BIBentrySTDinterwordspacing

\bibitem{HGS10}
\BIBentryALTinterwordspacing
P.~Hyllus, O.~G\"uhne, and A.~Smerzi, ``Not all pure entangled states are
  useful for sub-shot-noise interferometry,'' \emph{Phys. Rev. A}, vol.~82, p.
  012337, Jul 2010. [Online]. Available:
  \url{http://link.aps.org/doi/10.1103/PhysRevA.82.012337}
\BIBentrySTDinterwordspacing

\bibitem{Sa89}
\BIBentryALTinterwordspacing
B.~C. Sanders, ``Quantum dynamics of the nonlinear rotator and the effects of
  continual spin measurement,'' \emph{Phys. Rev. A}, vol.~40, pp. 2417--2427,
  Sep 1989. [Online]. Available:
  \url{http://link.aps.org/doi/10.1103/PhysRevA.40.2417}
\BIBentrySTDinterwordspacing

\bibitem{BIWH96}
\BIBentryALTinterwordspacing
J.~J.~. Bollinger, W.~M. Itano, D.~J. Wineland, and D.~J. Heinzen, ``Optimal
  frequency measurements with maximally correlated states,'' \emph{Phys. Rev.
  A}, vol.~54, pp. R4649--R4652, Dec 1996. [Online]. Available:
  \url{http://link.aps.org/doi/10.1103/PhysRevA.54.R4649}
\BIBentrySTDinterwordspacing

\bibitem{HO07}
\BIBentryALTinterwordspacing
H.~F. Hofmann and T.~Ono, ``High-photon-number path entanglement in the
  interference of spontaneously down-converted photon pairs with coherent laser
  light,'' \emph{Phys. Rev. A}, vol.~76, p. 031806, Sep 2007. [Online].
  Available: \url{http://link.aps.org/doi/10.1103/PhysRevA.76.031806}
\BIBentrySTDinterwordspacing

\bibitem{IARAS12}
\BIBentryALTinterwordspacing
Y.~Israel, I.~Afek, S.~Rosen, O.~Ambar, and Y.~Silberberg, ``Experimental
  tomography of noon states with large photon numbers,'' \emph{Phys. Rev. A},
  vol.~85, p. 022115, Feb 2012. [Online]. Available:
  \url{http://link.aps.org/doi/10.1103/PhysRevA.85.022115}
\BIBentrySTDinterwordspacing

\bibitem{PS08}
\BIBentryALTinterwordspacing
L.~Pezz\'e and A.~Smerzi, ``Mach-zehnder interferometry at the heisenberg limit
  with coherent and squeezed-vacuum light,'' \emph{Phys. Rev. Lett.}, vol. 100,
  p. 073601, Feb 2008. [Online]. Available:
  \url{http://link.aps.org/doi/10.1103/PhysRevLett.100.073601}
\BIBentrySTDinterwordspacing

\bibitem{BGB10}
\BIBentryALTinterwordspacing
G.~Brida, M.~Genovese, and R.~BercheraI., ``{Experimental realization of
  sub-shot-noise quantum imaging},'' \emph{Nat Photon}, vol.~4, no.~4, pp.
  227--230, Apr. 2010. [Online]. Available:
  \url{http://dx.doi.org/10.1038/nphoton.2010.29}
\BIBentrySTDinterwordspacing

\bibitem{TJDK13}
\BIBentryALTinterwordspacing
M.~A. Taylor, J.~Janousek, V.~Daria, J.~Knittel, B.~Hage, BachorHans-A., and
  W.~P. Bowen, ``{Biological measurement beyond the quantum limit},'' \emph{Nat
  Photon}, vol.~7, no.~3, pp. 229--233, Mar. 2013. [Online]. Available:
  \url{http://dx.doi.org/10.1038/nphoton.2012.346
  http://www.nature.com/nphoton/journal/v7/n3/abs/nphoton.2012.346.html\#supplementary-information}
\BIBentrySTDinterwordspacing

\bibitem{Tsa09}
\BIBentryALTinterwordspacing
M.~Tsang, ``Quantum imaging beyond the diffraction limit by optical centroid
  measurements,'' \emph{Phys. Rev. Lett.}, vol. 102, p. 253601, Jun 2009.
  [Online]. Available:
  \url{http://link.aps.org/doi/10.1103/PhysRevLett.102.253601}
\BIBentrySTDinterwordspacing

\bibitem{DJK14}
M.~J. R.~Demkowicz-Dobrzanski and J.~Kolodynski, ``{Q}uantum limits in optical
  interferometry,'' 2014, arXiv:1405.7703v2.

\bibitem{Kol14}
J.~Kolodynski, ``{P}recision bounds in noisy quantum metrology,'' Ph.D.
  dissertation, 2014, arXiv:1409.0535v1.

\bibitem{Hol82}
A.~S. Holevo, \emph{Probabilistic and Statistical Aspects of Quantum
  Theory}.\hskip 1em plus 0.5em minus 0.4em\relax North Holland, Amsterdam,
  1982.

\bibitem{JBWKLD12}
\BIBentryALTinterwordspacing
K.~Jiang, C.~J. Brignac, Y.~Weng, M.~B. Kim, H.~Lee, and J.~P. Dowling,
  ``Strategies for choosing path-entangled number states for optimal robust
  quantum-optical metrology in the presence of loss,'' \emph{Phys. Rev. A},
  vol.~86, p. 013826, Jul 2012. [Online]. Available:
  \url{http://link.aps.org/doi/10.1103/PhysRevA.86.013826}
\BIBentrySTDinterwordspacing

\bibitem{HWD08}
\BIBentryALTinterwordspacing
S.~D. Huver, C.~F. Wildfeuer, and J.~P. Dowling, ``Entangled fock states for
  robust quantum optical metrology, imaging, and sensing,'' \emph{Phys. Rev.
  A}, vol.~78, p. 063828, Dec 2008. [Online]. Available:
  \url{http://link.aps.org/doi/10.1103/PhysRevA.78.063828}
\BIBentrySTDinterwordspacing

\bibitem{RJD13}
\BIBentryALTinterwordspacing
B.~Roy~Bardhan, K.~Jiang, and J.~P. Dowling, ``Effects of phase fluctuations on
  phase sensitivity and visibility of path-entangled photon fock states,''
  \emph{Phys. Rev. A}, vol.~88, p. 023857, Aug 2013. [Online]. Available:
  \url{http://link.aps.org/doi/10.1103/PhysRevA.88.023857}
\BIBentrySTDinterwordspacing

\end{thebibliography}

\end{document}